\documentclass[10pt,pra,letterpaper,typeset,twocolumn,superscriptaddress,showpacs,floatfix]{revtex4-1}
\usepackage{color}
\usepackage{graphicx}
\usepackage{amsmath}
\usepackage{amssymb,amsthm}
\graphicspath{{pict/}{}}

\usepackage{pgf,tikz}
\usepackage{mathrsfs}
\usepackage{bm}
\usepackage{array}
\usepackage{blkarray}

\usepackage{hyperref}
\hypersetup{colorlinks=true,linkcolor=blue,citecolor=blue,urlcolor=blue}

\newcounter{Fig}

 \newcommand{\ket}[1]{\left|#1\right\rangle}
 \newcommand{\bra}[1]{\left\langle#1\right|}

\begin{document}

\title{Quantum localized states in photonic flat-band lattices}

\author{S. Rojas-Rojas}
\affiliation{Center for Optics and Photonics and MSI-Nucleus on Advanced Optics, Universidad de Concepci\'{o}n, Casilla 160-C, Concepci\'{o}n, Chile}
\affiliation{Departamento de F\'{i}sica, Universidad de Concepci\'{o}n, Casilla 160-C, Concepci\'{o}n, Chile}
 
\author{L. Morales-Inostroza}
\affiliation{Center for Optics and Photonics and MSI-Nucleus on Advanced Optics, Universidad de Concepci\'{o}n, Casilla 160-C, Concepci\'{o}n, Chile}
\affiliation{Departamento de F\'isica, Facultad de Ciencias, Universidad de Chile, Santiago, Chile}

\author{R. A. Vicencio}
\affiliation{Center for Optics and Photonics and MSI-Nucleus on Advanced Optics, Universidad de Concepci\'{o}n, Casilla 160-C, Concepci\'{o}n, Chile}
\affiliation{Departamento de F\'isica, Facultad de Ciencias, Universidad de Chile, Santiago, Chile} 
 
\author{A. Delgado}
\affiliation{Center for Optics and Photonics and MSI-Nucleus on Advanced Optics, Universidad de Concepci\'{o}n, Casilla 160-C, Concepci\'{o}n, Chile}
\affiliation{Departamento de F\'{i}sica, Universidad de Concepci\'{o}n, Casilla 160-C, Concepci\'{o}n, Chile}


\begin{abstract}

The localization of light in flat-band lattices has been recently proposed and experimentally demonstrated in several configurations, assuming a classical description of light. Here, we study the problem of light localization in the quantum regime. We focus on quasi one-dimensional and two-dimensional lattices which exhibit a perfect flat-band inside their linear spectrum. Localized quantum states are constructed as eigenstates of the interaction Hamiltonian with a  vanishing eigenvalue and a well defined total photon number. These are superpositions of Fock states with probability amplitudes given by positive as well as negative square roots of multinomial coefficients. The classical picture can be recovered by considering poissonian superpositions of localized quantum states with different total photon number. We also study the separability properties of flat band quantum states and apply them to the transmission of information via multi-core fibers, where these states allow for the total passive suppression of photon crosstalk and exhibit robustness against photon losses. At the end, we propose a novel on-chip setup for the experimental preparation of localized quantum states of light for any number of photons.

\end{abstract}

\maketitle

\section{Introduction}

Localization of light in extended periodical and homogeneous lattices may occur due to different causes. For example, the interplay between discreteness and nonlinearity --- or equivalently, between diffraction and self-focusing --- in a perfectly periodic lattice leads to the generation of discrete solitons (or discrete breathers)~\cite{Flach,rep1}; that is, spatially localized stationary nonlinear modes~\cite{Cambpell}. On the other hand, Anderson localization~\cite{Anderson} is a linear phenomena appearing in disordered lattices, where consecutive destructive interference from randomly distributed scatters (lattice sites) suppresses the transversal transport across the lattice, generating localized discrete states~\cite{segev,morandotti}. 

However, a new mechanism for the emergence of localization in discrete periodic and linear lattices has been theoretically suggested recently~\cite{berg1}. This effect arises in specific lattice configurations, which allow the  cancellation of amplitudes at different sites of the lattice, generating localized stationary states in the absence of disorder or nonlinearity. Certain lattices exhibit a linear energy spectrum composed of at least one dispersive band and a perfectly flat band (FB), which has a large set of degenerated localized states. Some recently investigated quasi one-dimensional (1D) and two-dimensional (2D) geometries include: \emph{rhomboidal}~\cite{sebarom}, \emph{stub}~\cite{Amo}, \emph{sawtooth}~\cite{uta,chi1,Weimann}, \emph{kagome-ribbon}~\cite{kagrib1}, \emph{Lieb}~\cite{njpLieb,Vicencio1,Mukherjee1,chenLieb,anton2} and Kagome~\cite{stan1,Vicencio2,chenKag} lattices. FB modes are spatially trapped due to the destructive interference induced by the specific lattice geometry~\cite{Luis}, and remain localized due to the absence of dispersion. Thereby, localization is exact and independent of external parameters or extra interactions. The study of flat bands lattices initially arises in the context of condensed matter, where many complex phenomena are difficult to observe directly in an experiment. For this reason, simpler physical models with the same geometry and band structure have been proposed as simulators~\cite{polini1}, being photonics lattices or waveguide arrays one of the most suitable systems for the observation of discrete phenomena in any physical context~\cite{Flach,rep1}.

\begin{figure*}[!tb]
	\includegraphics[width=.9\textwidth]{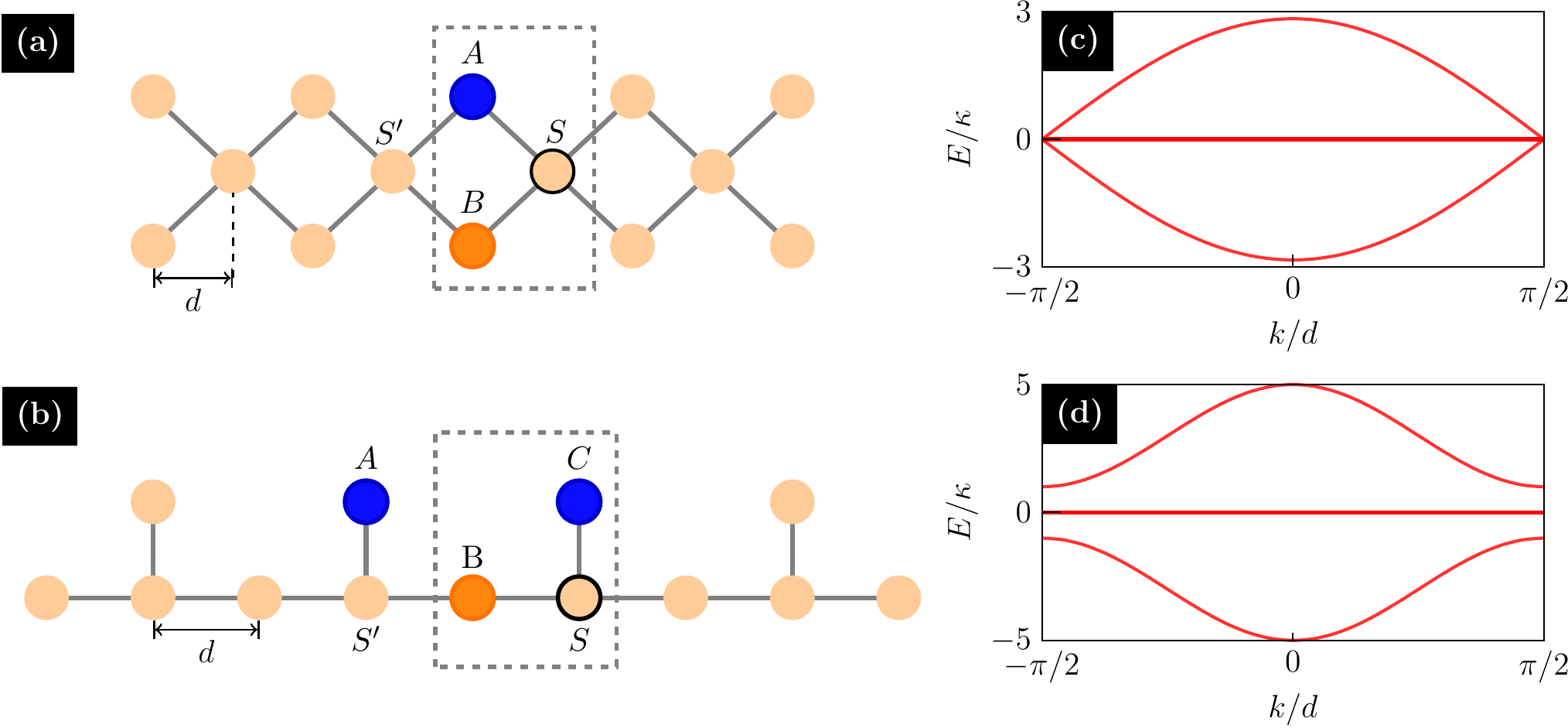}
	\caption{(a) Rhomboidal and (b) stub lattices. Gray lines depict the evanescent coupling between sites. We highlight certain sites which compose a localized state of the flat-band, and the connectors $S$ with the rest of the lattice. Dashed rectangles indicate the unitary cell of the lattice. (c) and (d) Frequency spectrum of the rhomboidal and stub lattices, respectively.}\label{fg:geom}
\end{figure*}

To our knowledge, previous works on flat-band photonic lattices have only considered a classical description of light fields. On the other hand, many studies  have addressed the quantum properties of light propagating along unidimensional waveguide arrays (which exhibit dispersive bands only). It has been shown that the propagation of single photons in these lattices provides an implementation of quantum walks \cite{quwalks}, while the propagation of photon pairs leads to nontrivial quantum correlations which depend critically on the input state \cite{morandotti2,Peruzzo}. Furthermore, the persistence of such correlations in strongly disordered systems has been observed \cite{epr}. Here, we aim to expand these results by studying the existence and properties of localized (non-diffractive) quantum states of light on flat-band lattices, for an arbitrary number of photons. This is mainly motivated by the possibility of employing localized photonic states for the secure transmission of information and entanglement via multi-core fibers (MCF)~\cite{Canas,Ding}. 

We initially focus on the rhomboidal (diamond) and stub lattices, and consider quantum states with a well defined total number of photons. The general definition of the  minimal localized eigenstates is derived, which turns out to be a finite superposition of multipartite Fock states weighted by real (positive and negative) probability amplitudes given by the square root of multinomial coefficients. The corresponding definitions for other common flat-band lattices is presented too.  We show that the quantum description of the localized classical states of light corresponds to poissonian superpositions of localized quantum states with different total number of photons. We also study the separability properties of the localized states. In the rhomboidal lattice, all the localized eigenstates are composed of Fock states corresponding to two lattice sites only, and are entangled for any total number of photons. In the stub lattice, flat-band eigenstates extend over three sites. We show that any bipartite reduced state (obtained after tracing out a site) is entangled, implying pairwise entanglement equally distributed over the sites. In this case, the single-photon localized state is a W-state. For higher dimensions, however, we find that the localized eigenstates do not satisfy the monogamy relation characteristic of generalized W-states~\cite{KyS}. At the end, we apply our results to the propagation of light in a four-core MCF. In this context, photon losses and inter-core crosstalk are important sources of errors. These can be effectively corrected with the help of the non-diffractive states introduced here. Crosstalk is passively suppressed for any coupling constant between nearest neighbors. Photon losses reduce the total number of photons without either destroying the localization properties nor transforming the states into separable ones. Thereby, the propagation of entanglement via MCF seems possible. Finally, we present a novel method to prepare localized quantum states in flat-band photonic chips. 

\section{Model}

We consider the propagation of light in an array of evanescently-coupled identical waveguides. In the classical case, where the light injected into the array is a coherent state, with a large average photon number, the dynamics of the wave packet propagating along the waveguides is well described by Discrete Linear Schr\"odinger (DLS) equation~\cite{Flach,rep1}. In this context, every waveguide corresponds to a node or site in a given lattice, and the coupling of light between neighboring sites is weak due to its evanescent nature. On the other hand, in the quantum case the propagation of light can be described accurately by means of the following Hamiltonian operator
\begin{equation}\label{eq:H}
H=\sum_n\epsilon_na^\dagger_na_n-\sum\limits_{n,m} \left(\kappa_{n,m}a^\dagger_n a_m+\kappa_{m,n}a^\dagger_n a_m\right)\ .
\end{equation}
Here $a^\dagger_n$ and $a_n$ correspond to the creation and the annihilation operators for the mode at the $n$-th waveguide, respectively, while $\epsilon_n$ describes the longitudinal propagation constant at waveguide $n$. The coupling interaction between nearest-neighbors waveguides is included in this model through the second term at the right hand side of Eq.\thinspace\eqref{eq:H}, which resembles the tight-binding model from solid-state physics~\cite{kittel}. Here, $\kappa_{n,m}$ denotes the coupling constant for photon hopping between waveguides $n$ and $m$. In what follows, states of light will be spanned as linear combinations of tensor products of Fock states $|p\rangle_m$, which describes $p$ photons at waveguide $m$. In the limit corresponding to a large number of photons per site the evolution under the Hamiltonian $H$, Eq.\thinspace\eqref{eq:H}, it is well approximated by the DLS models~\cite{Scott,Pinto}.

Depending on the transverse dispositions of the waveguides, it is possible to obtain a periodic array whose spectrum of eigenfrequencies exhibits a nondispersive flat band; that is, a subset of degenerated states, which are perfectly localized in space (with a zero tail, similar to compactons~\cite{comp}). There are many different geometries which present at least one FB.  Along this work, we will consider two quasi one-dimensional arrays with perfect flat bands: (i) the rhomboidal lattice, to this date the simplest lattice presenting a flat band; and (ii) the stub lattice, which is useful to clarify the way to extend our results to more complex geometries. These two lattices are illustrated in Fig.\thinspace\ref{fg:geom}. In both cases we assume homogeneous lattices with identical waveguides for which $\epsilon_n=\epsilon$. We also consider a symmetric coupling interaction between waveguides, that is, $\kappa_{n,m}=\kappa_{m,n}=\kappa$. Thereby, the Hamiltonian operator reads
\begin{equation}
H=H_0+H_{int}\ ,
\label{Hamiltonian}
\end{equation}
where the free evolution Hamiltonian $H_0$ operator is given by
\begin{equation}
H_0=\epsilon\sum_na^\dagger_n a_n\ ,
\label{FreeHamiltonian}
\end{equation}
and the interaction Hamiltonian operator $H_{int}$ is
\begin{equation}
H_{int}=-\kappa\sum\limits_{n,m}(a^\dagger_n a_m+a^\dagger_m a_n)\ .
\label{InteractionHamiltonian}
\end{equation}

\section{Localized quantum states of light}

A photonic lattice with a quasi 1D rhomboidal lattice has a unitary cell which contains only three lattice sites: $A$, $B$ and $S$, as shown in Fig.\thinspace\ref{fg:geom}(a). Therefore, in the classical regime, the rhomboidal lattice possesses only three energy bands, where one of them is perfectly flat (non-dispersive). Within this band it is possible to generate a superposition of degenerated eigenmodes, located in different regions of the lattice, which exhibits a difractionless propagation due to the zero dispersion of this FB. These localized states are generated by injecting coherent light in two waveguides, $A$ and $B$ as shown in Fig.\thinspace\ref{fg:geom}(a), with equal amplitude but opposite phase ($\pi$)~\cite{sebarom}. This phase difference plays an important role: it leads to destructive interference between evanescent waves originated at waveguides $A$ and $B$, cancelling the coupling of energy to the central row and, thus, keeping the light perfectly localized in space.

States produced in this way are degenerated eigenmodes of the DLS equation with zero frequency. This suggests, in the quantum case, to look for states which are annihilated by the action of the interaction Hamiltonian operator. Let us consider for instance the single photon state $\alpha|0\rangle_A|1\rangle_B+\beta|1\rangle_A|0\rangle_B$, where the resting waveguides are in the vacuum state. The action of the interaction Hamiltonian operator Eq.\thinspace(\ref{InteractionHamiltonian}) onto this state generates the new state $(\alpha+\beta)(|1\rangle_{S'}|0\rangle_{S}+|0\rangle_{S'}|1\rangle_{S})|0\rangle_A|0\rangle_B$, which vanishes when $\alpha$ and $\beta$ differ by a phase of $\pi$. This phase difference cancels the hopping of the photon from waveguides $A$ and $B$ towards waveguides $S$ and $S'$, in an analogous manner to the classical case, since the not normalized state $|0\rangle_A|1\rangle_B-|1\rangle_A|0\rangle_B$ is annihilated by the operators $a_{S}^\dagger a_A+a_{S}^\dagger a_B$ and $a_{S}^\dagger a_A+a_{S'}^\dagger a_B$.

The previous considerations can be readily extended to the case of states with a well defined total number $N\ge1$ of photons. We start with the state
\begin{equation}
|\psi\rangle_{A,B}=\sum\limits^{N}_{p,q}C_{p,q}|p\rangle_A|q\rangle_B\ ,
\end{equation}
where all other waveguides are initially in the vacuum state and the summation is carried out under the condition $p+q=N$.  This state is an eigenstate of the total photon number $a^\dagger_Aa_A+a^\dagger_Ba_B$ with eigenvalue $N$ for any set $\{C_{p,q}\}$ of probability amplitudes. Since most of the waveguides are initially in the vacuum state, the action of the interaction Hamiltonian operator involves waveguides $A$, $B$, $S$ and $S'$ only. In this sector, the interaction Hamiltonian can be split into two contributions $H_{S}=a_{S}^\dagger a_A+a_{S}^\dagger a_B+h.c.$ and $H_{S'}=a_{S'}^\dagger a_A+a_{S'}^\dagger a_B+h.c.$, which describe the hopping of a single photon between waveguides $A$ or $B$ and waveguides $S$ and $S'$, respectively. Since the initial state $|\psi\rangle$ must stay localized in waveguides $A$ and $B$ along the evolution, photon hopping toward sites $S$ and $S'$ cannot occur. To assure this, we impose the condition $H_{S}|\psi\rangle_{A,B}=H_{S'}|\psi\rangle_{A,B}=0$, that is, the state is annihilated by both operators $H_{S}$ and $H_{S'}$. Due to the symmetry of the lattice, both operators can be considered independently, so it is enough to evaluate the action of $H_{S}$ onto $|\psi\rangle_{A,B}$, obtaining the state
\begin{equation}
\sum^{N-1}_{p,q}(C_{p+1,q}\sqrt{p+1}+C_{p,q+1}\sqrt{q+1})|1\rangle_{S}|p\rangle_A|q\rangle_B\,,
\end{equation}
which together with the condition $H_{S}|\psi\rangle_{A,B}=0$ allows us to find the following recursive relation:
\begin{equation}
C_{p+1,q}=-\frac{\sqrt{q+1}}{\sqrt{p+1}}C_{p,q+1}\ .
\end{equation}
This and the normalization condition, $\sum\limits^{N}|C_{p,q}|^2=1$, lead to the solution
\begin{equation}
C_{p,q}=(-1)^q\binom{N}{p\, q}^{1/2}\frac{1}{2^{N/2}}\ ,
\label{Cs}
\end{equation}
where $\binom{N}{p,q}=N!/p!\,q!$ is the binomial coefficient.
Thereby, the localized quantum state of light with $N$ photons in the rhomboidal lattice is given by
\begin{equation}\label{eq:psinr}
\ket{\psi_N}^r_{A,B}=2^{-\frac{N}{2}}\sum\limits^{N }_{p,q}\!
\binom{N}{p\,q}^{\frac{1}{2}}\,(-1)^{q}\ket{p}_A\ket{q}_B\ .
\end{equation}
Light described by this state propagates along waveguides $A$ and $B$ without diffracting to the neighboring waveguides $S'$ and $S$, that is, it stays perfectly localized for any number $N$ of photons and independently of the value of the coupling constant $\kappa$. Additionally, each unit cell in the rhomboidal lattice admits the existence of this class of states. Let us note that $\ket{\psi_N}^r_{A,B}$ is also annihilated by  the operator $H_{+1}$ and thus it is annihilated by the interaction Hamiltonian. States $|\psi_N\rangle^r_{A,B}$ are mutually orthogonal; that is, $_{A,B}\thinspace^r\langle\psi_{N}|\psi_{N'}\rangle^r_{A,B}=\delta_{N,N'}$, and can be, perfectly and deterministically, distinguished through a measurement of the total photon number.

Our construction of localized states for the rhomboidal lattice can be readily extended to the stub geometry [see Fig.\thinspace\ref{fg:geom}(b)]. In this case, the classical localized state corresponds to coherent light distributed with equal intensity over three waveguides, say $A$, $B$ and $C$, as depicted in Fig.\thinspace\ref{fg:geom}(b), with a phase difference of $\pi$ between waveguides $A$ and $B$ and waveguides $B$ and $C$~\cite{Amo}. Thereby, in close analogy to the rhomboidal lattice, we construct a quantum state with $N$ photons, which is annihilated by the operators $a_{S'}^\dagger a_A+a_{S'}^\dagger a_B$ and $a_{S}^\dagger a_B+a_{S}^\dagger a_C$. We find that the localized quantum state of light for the stub lattice is
\begin{equation}\label{eq:psins}
\ket{\psi_N}^s_{A,B,C}= 3^{-\frac{N}{2}}\sum\limits^{N }_{p,q}
\binom{N}{p\,q\,t}^{\frac{1}{2}}(-1)^q\ket{p}_{A}\ket{q}_{B}\ket{t}_{C}.
\end{equation}
Analogously, localized quantum states of light for Lieb, symmetric rhomboidal, and Kagome lattices are presented in Table \ref{tab:states}.
\begin{table*}[!bt]
	\begin{tabular}{m{2.2cm} m{2.5cm} m{3cm} m{9cm}}
				Lattice & ND-state sites & Flat band frequency & Eigenstate \\
		\tikz \node[text width=2cm] {Lieb}; &
		\vspace*{.25em}\includegraphics[width=1.8cm]{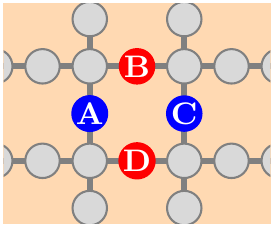}  & $0$ & ${\displaystyle \frac{1}{4^{N/2}}\sum\limits_{\begin{smallmatrix}p,q,t,\ell\geq 0\\ p+q+t+\ell=N 
				\end{smallmatrix}}\!
			\binom{N}{p\,q\,t\,\ell}^{\frac{1}{2}}\,(-1)^{q+\ell}\ket{p}_{A}\ket{q}_{B}\ket{t}_{C}\ket{\ell}_{D}}$ \\ \hline  
		\tikz \node[text width=2cm] {Symmetric Rhomboidal}; & 
		\vspace*{.25em}\includegraphics[width=1.8cm]{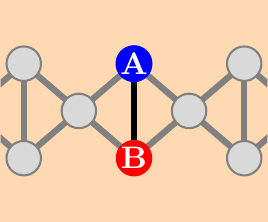} & $\kappa N$ & ${\displaystyle \frac{1}{2^{N/2}}\sum\limits_{\begin{smallmatrix}p,q\geq 0\\ p+q=N 
				\end{smallmatrix}}\!
			\binom{N}{p\,q}^{\frac{1}{2}}\,(-1)^{q}\ket{p}_A\ket{q}_B}$\\ \hline
		\tikz \node[text width=2cm] {Kagome}; & 
		\vspace*{.25em}\includegraphics[width=1.8cm]{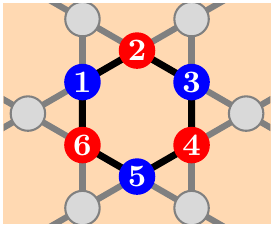} & $2\kappa N$ & ${\displaystyle \frac{1}{6^{N/2}}\sum\limits_{\begin{smallmatrix}p_1,p_2\ldots p_6\geq 0\\ p_1+p_2+\cdots+p_6=N 
				\end{smallmatrix}}\!
			\binom{N}{p_1,p_2,\ldots,p_6}^\frac{1}{2}\,(-1)^{p_2+p_4+p_6}\bigotimes\limits_{i=1}^6\ket{p_i}_i}$
	\end{tabular}
 \caption{Localized quantum states for some standard flat-band lattices.}
 \label{tab:states}
\end{table*}

The localized states we have constructed are annihilated by the interaction Hamiltonian. Thereby, only the free Hamiltonian $H_0$, Eq.\thinspace\eqref{FreeHamiltonian}, will determine their evolution along $z$ (the dynamical variable in our system). Namely, the action of the evolution operator onto these states is simply given by
\begin{equation}
U(z)\ket{\psi_N}=e^{-i\epsilon Nz}\ket{\psi_N}\ ,
\end{equation}
where $\ket{\psi_N}$ is a localized state with $N$ photons corresponding to a fixed unitary cell, for any of the FB lattices previously studied (notice that it holds true for lattices whose FB is not located at zero frequency, which only modifies the effective value of the propagation constant $\epsilon$). Any linear superposition of localized quantum states, such as $\ket{\psi}=\sum_{N=0}^{\infty}D_N\ket{\psi_N}$, will evolve as
\begin{equation}
U(z)|\psi\rangle=\sum_{N=0}^{\infty}D_Ne^{-i\epsilon Nz}|\psi_N\rangle\ .
\end{equation}
This superposition remains localized at the same sites of states $\ket{\psi_N}$, for any set $\{D_N\}$ of coefficients. The state, however, changes along the evolution due to the phase factors $e^{i\epsilon Nz}$.

A particularly interesting superposition of localized states is given by
\begin{equation}
\ket{\beta}^r_{A,B}=e^{-|\beta|^2/2}\sum\limits_{N=0}^\infty \frac{\beta^N}{\sqrt{N!}}\ket{\psi_N}^r_{A,B}\ ,
\end{equation}
which corresponds to a poissonian superposition of localized states $\ket{\psi_N}^r_{A,B}$. Considering the case $\beta=\sqrt{2}\alpha$, the previous state becomes
\begin{equation}
|\sqrt{2}\alpha\rangle^r_{A,B}=\ket{\alpha}_{A}\ket{-\alpha}_B,
\end{equation}
that is, a tensor product of a coherent state for each waveguide $A$ and $B$ with a phase difference of $\pi$ between them. This state is the quantum analogue of the classical localized state, which is generated by injecting coherent light of equal intensity on each waveguide with a phase difference of $\pi$. Thus, the quantum description of the classical localized state corresponds to a poissonian distribution onto localized quantum states $|\psi_N\rangle^r_{A,B}$.

So far we have considered that in the rhomboidal lattice photon hopping between waveguides $A$ and $B$ does not occur. Nevertheless, even in presence of 
 coupling between these waveguides, which is the case for the symmetric rhomboidal lattice (see table \ref{tab:states}), states $|\psi_N\rangle^r_{A,B}$ remain localized and do not evolve. This is due to the fact that these states are also eigenstates of the interaction operator $a^\dagger_Aa_B+a_Aa^\dagger_B$ with an eigenvalue proportional to $-N$. The existence of this coupling preserves the flat band structure up to a shift of the value of $\epsilon$. Unlike the case of the rhomboidal lattice, stub, Lieb and Kagome lattices do not exhibit flat bands whenever next nearest-neighbor interactions are considered.

\section{Entanglement of localized quantum states of light} 

The quantum description of classical localized light corresponds to a tensor product of coherent states. Consequently, this state is clearly unentangled. As we shall see, this is not the case for the $N$-photon localized quantum states. 

\subsection{Rhomboidal lattice}

Let us start by considering the simplest case of the single-photon localized state of the rhomboidal lattice,
\begin{equation}
\ket{\psi_1}^r_{A,B}=\frac{1}{\sqrt{2}}\left(\ket{1}_{A}\ket{0}_{B}-\ket{0}_{A}\ket{1}_{B}\right)\,,
\end{equation}
which corresponds to a path-entangled state~\cite{Jones,Enk}. In the general case, since the $N$-photon states $\ket{\psi_N}^r$ are pure, whether they are entangled or not can be determined directly from their Schmidt decomposition. From Eq.\thinspace\eqref{eq:psinr} we obtain that the $i$-th Schmidt coefficient of state $\ket{\psi_N}^r$ is given by
\begin{equation}\label{eq:schmcfr}
 k_{i,N}=\frac{1}{2^{N/2}}\binom{N}{i}^{1/2}\ ,
\end{equation}
with $i=0,\dots,N$. Thus, each state $\ket{\psi_N}^r$ has $N+1$ non vanishing Schmidt coefficients.

\begin{figure}[!b]
 \centering
 \includegraphics[width=.47\textwidth]{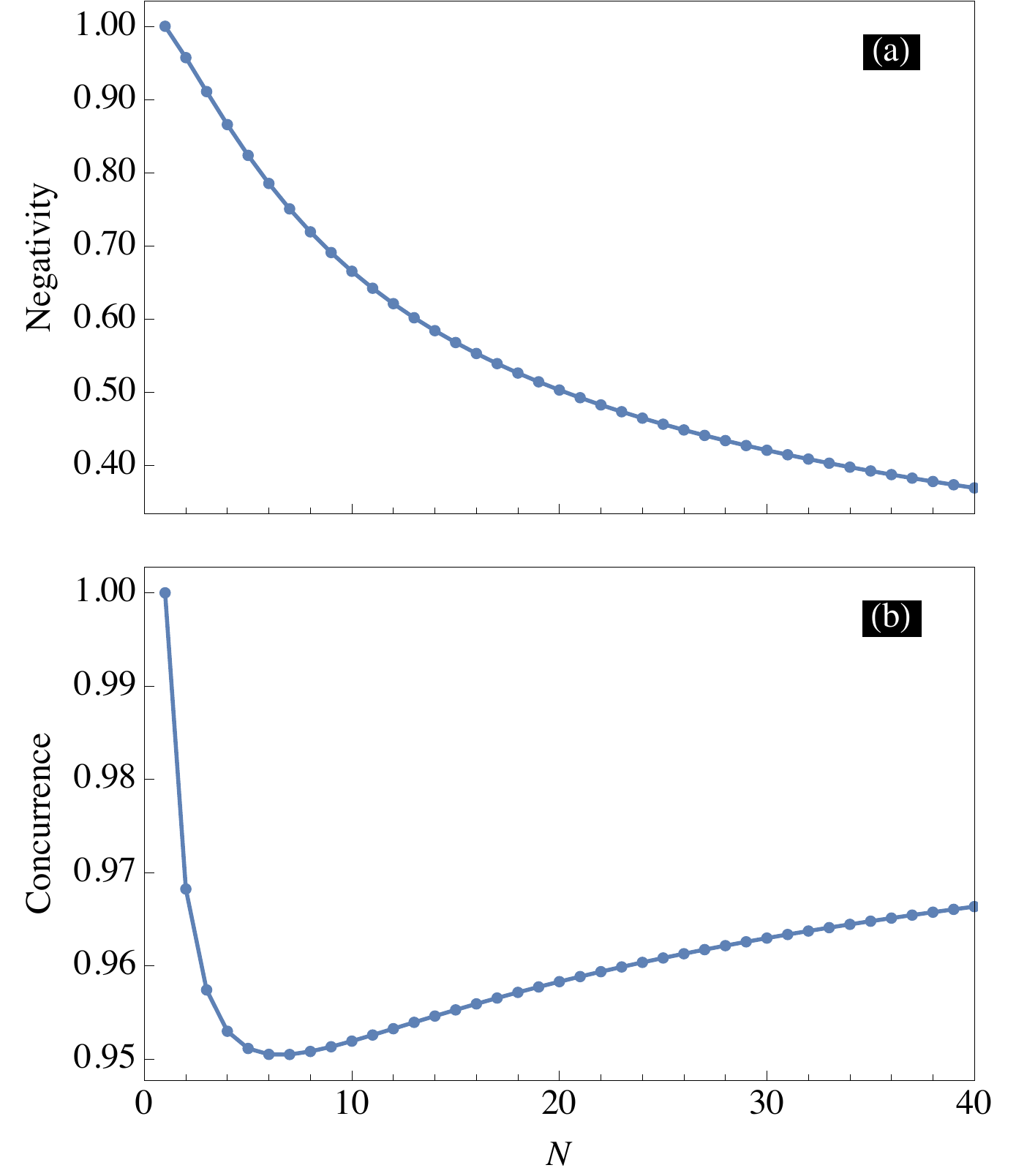}
 \caption{(a) Negativity and (b) concurrence of the non-diffractive photon states of the rhomboidal lattice, as a function of the photon number $N$.}\label{fg:ent}
\end{figure}

The Schmidt decomposition is closely linked to two measures of entanglement, namely, {\emph{negativity}} and \emph{concurrence}~\cite{neg,ent}. The former is given by $\mathcal{N}(\ket{\psi_N}^r_{A,B})=(\|\rho^{T_A}_N\|-1)/(d-1)$, where $\rho^{T_A}_N$ is the partial transpose (PT) of the density matrix $\rho_N=|\psi_N\rangle^r\langle\psi_N|$ with respect to waveguide $A$, and $d$ is the dimension of each subsystem, which in our case is $d=N+1$. The value of negativity for a separable state is zero, while for a maximally entangled state it is equal to one~\cite{Lee}. Negativity of a localized eigenstate $\ket{\psi_N}^r_{A,B}$ can be expressed in terms of its Schmidt coefficients Eq.\thinspace\eqref{eq:schmcfr} as
\begin{eqnarray}
 \mathcal{N}(\ket{\psi_N}^r_{A,B})&=&\frac{2}{N}\sum\limits_{i<j}k_{i,N}k_{j,N}\ ,
 \nonumber\\
 &=&\frac{1}{2^N N}\left(\sum_{i=0}^N\binom{N}{i}^{1/2}\right)^2-\frac{1}{N}\ .
\label{eq:negk}
\end{eqnarray}
As illustrated in Fig.\thinspace\ref{fg:ent}(a), negativity is positive for any state $\ket{\psi_N}^r_{A,B}$ with a  finite number $N$ of photons, but decreases monotonically as $N$ grows. In the asymptotic limit $N\longrightarrow \infty$, negativity decays to zero as $\sqrt{2/\pi}N^{-3/2}-1/N$.  As far as it is not null, the corresponding state $\ket{\psi_N}^r_{A,B}$ is entangled.

Concurrence is defined as $\sqrt{2(1-{\rm Tr}\rho_A^2)}/\mathcal{C}^{\rm max}_n$, where $\rho_A$ is the reduced density matrix 
obtained after tracing the site $B$ from the full bipartite state. In our definition, we include the factor $1/{\mathcal C}^{\rm max}_N$, in order to express the concurrence relative to its maximum value on each dimension, namely, ${\mathcal C}^{\rm max}_N=\sqrt{2[1-1/(N+1)]}$. Both negativity and concurrence of the state $\ket{\psi_1}^r$ are equal to $1$, since it is a maximally entangled state. Concurrence, like negativity, can be expressed in terms of the Schmidt coefficients. This way, concurrence of the $\ket{\psi_N}^r$ state reads
\begin{align}
\mathcal{C}(\ket{\psi})&=\frac{2}{{\mathcal C}^{\rm max}_N}\left(\sum\limits_{i<j}k_{i,N}^2k_{j,N}^2\right)^{1/2}\ ,
\nonumber\\
 &=\frac{2^{1/2-N}}{{\mathcal C}^{\rm max}_N}\sqrt{2^{2N}-\binom{2N}{N}}\ .
\label{eq:cck}
\end{align}
Fig.\thinspace\ref{fg:ent}(b) shows the concurrence of states $\ket{\psi_N}^r$. A pure state with non-zero concurrence is entangled. We observe that as the number of photons increases, the corresponding concurrence exhibits a slight decrease, after which it slowly tends to its maximum value. For $N=10^3$ it differs from the unity by less than $10^{-3}$. This can be checked directly from Eq.\thinspace (\ref{eq:cck}), by evaluating the limit $N\longrightarrow \infty$ and noting that concurrence then takes the asymptotic value $\sqrt{2}/{\mathcal C}^{\rm max}_N\longrightarrow 1$.

\subsection{Stub lattice}

In the case of the stub lattice, the localized eigenstates  involve three sites, so their entanglement properties are difficult to quantify. Althought both negativity and concurrence can be defined for this case, measurement of these quantities is not conclusive, since entangled states with no concurrence or tripartite negativity have been found~\cite{wein}. Also, resort to the Schmidt decomposition is not a simple choice, because there are many different ways to extend its definition to multipartite systems~\cite{acin,eisert,carteret,sokoli}. For instance, if we try to express a state $\ket{\psi_N}^s_{A,B,C}$ in  the general decomposed form introduced by Carteret {\it et.\thinspace al.}~\cite{carteret}, we find that it is not possible by means of local transformations only. On the other hand, the intuitive (but more restrictive) generalization proposed by Sokoli and Alber \cite{sokoli} can not be applied to our states $\ket{\psi_N}^s_{A,B,C}$, since these do not fulfill the conditions to be \emph{Schmidt decomposable}. Indeed, the single-photon localized state $\ket{\psi_1}^s_{A,B,C}=(\ket{1}_A-\ket{1}_B+\ket{1}_C)/\sqrt{3}$ is entangled, and it corresponds to a W-state. For this kind of states, not even a general normal form~\cite{normal} can be constructed.

Even without the Schmidt decomposition, we can show that bipartite partitions of photons propagating along waveguides at sites A, B and C in the states $\ket{\psi_N}^s_{A,B,C}$ are entangled superpositions. For this purpose, we define the base states
\begin{equation}\label{eq:ip}
 \ket{i'_N}_{B,C}=\sum_{\mu\nu}A_{i\mu\nu}^{(N)}\ket{\mu}_B\ket{\nu}_C\,,
\end{equation}
where $A_{pqt}^{(N)}$ is the respective coefficient accompanying the state $\ket{p}_A\ket{q}_B\ket{t}_C$ in definition \eqref{eq:psins} for the state $\ket{\psi_N}^s_{A,B,C}$. Note that states $\ket{i'_N}_{B,C}$ satisfy the recurrence relation $\ket{i'_N}_{B,C}=\ket{(i-1)'_{N-1}}_{B,C}$. 

Then, the state $\ket{\psi_N}^s_{A,B,C}$ can be expressed in the form
 \begin{equation}\label{eq:sdecom}
  \ket{\psi_N}^s_{A,B,C}=\sum_i K_{i,N}\ket{i}_{A}\ket{i'_N}_{BC}\,,
 \end{equation}
with coefficients
\begin{equation}\label{eq:schmcfs}
 K_{i,N}=\frac{1}{3^{N/2}}\binom{N}{i,N-i}^{1/2}\left[\begin{tikzpicture}[baseline=(current bounding box.center)] \node[anchor=base,inner sep=0pt]{$\displaystyle \sum_{\begin{smallmatrix}p,q,t\geq 0\\ p+q+t=N-i 
\end{smallmatrix}}$};\end{tikzpicture}\!
\binom{N}{p,q,t}\right]^{1/2}\,,
\end{equation}
for $i=0,1,\ldots,N$. We remark that, in general, a tripartite Schmidt decomposition in the form of Eq.\thinspace\eqref{eq:sdecom} \emph{is not possible}~\cite{base} and from the definition of states $\ket{i'_N}_{BC}$ we see that in order to construct our decomposition it is neccesary to operate jointly on sites B y C, i.\thinspace e., a nonlocal operation is required, contrary to the Schmidt decomposition method. Anyway, since the coefficients in Eq.~(\ref{eq:schmcfs}) are not null for $i=0,1,\ldots,N$, we see that decomposition of Eq.~(\ref{eq:sdecom}) has $N+1$ different non-factorable terms, suggesting that the states $\ket{\psi_N}^s_{A,B,C}$ are bipartite entangled.

\begin{figure}[!tb]
	\centering
	\includegraphics[width=.48\textwidth]{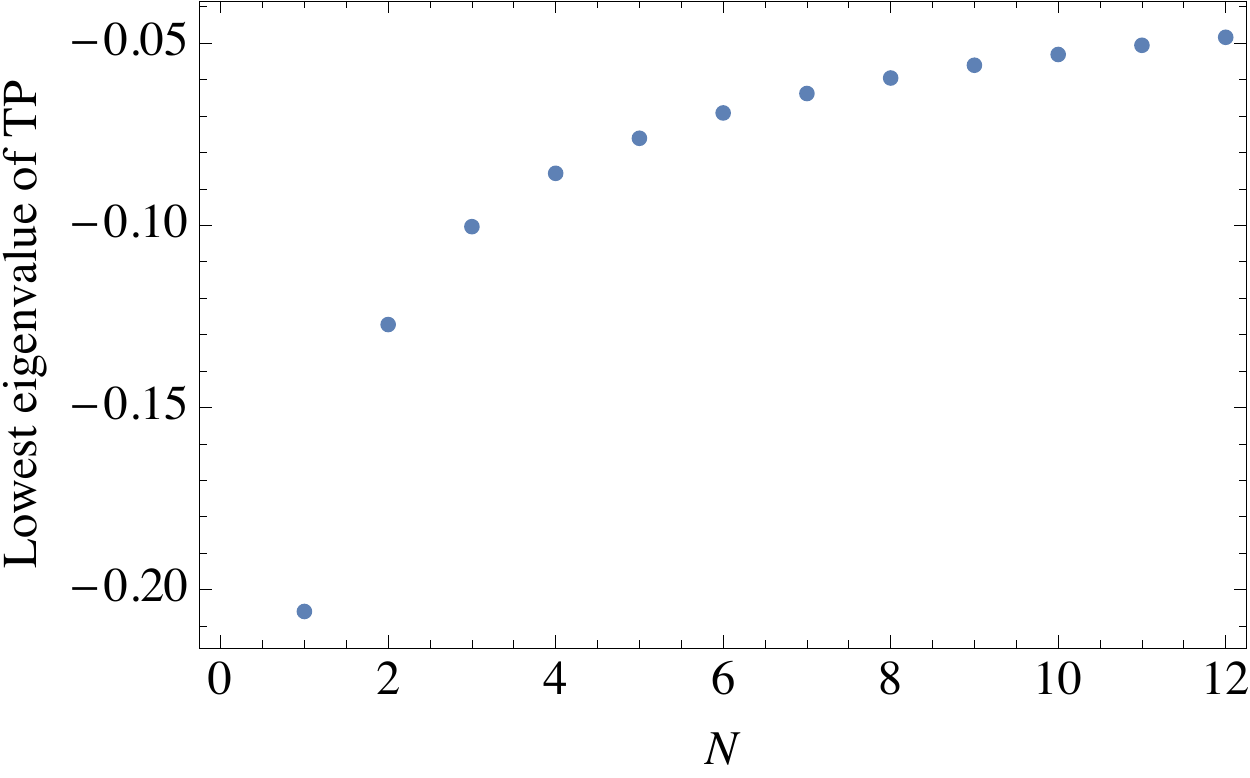}
	\caption{Minimum eigenvalue of the partial transposed reduced matrix obtained by remotion of any party in states $\ket{\psi_N}^s$, for different numbers of photons. The existence of a negative eigenvalue indicates the entanglement of the reduced states.}\label{fg:redneg}
\end{figure}

In order to confirm the previous observation, we study the bipartite states obtained by removing (tracing) one of the parties composing the full state $\ket{\psi_N}^s$ (this can be the effect of a measurement in one of the waveguides). Let $M_N$ be the partial transpose of the density matrix corresponding to the reduced state. Then, according to the Peres-Horodecki (PH) criterion~\cite{Peres,Horodecki}, if $M_N$ has at least one negative eigenvalue, the reduced state will be entangled. In Fig.\thinspace \ref{fg:redneg} we show the minimum eigenvalue of $M_N$ computed up to $N=12$ photons. The results are the same, whichever the deleted party be. We observe that there is a negative eigenvalue for all the numbers of photons considered. Since this holds for any choice of the sites in the reduced state, we conclude that states $\ket{\psi_N}^s$ are pairwise-entangled, and its entanglement is equally distributed over all the possible pairs of paths.  The fact that entanglement remains after removing a party, suggests that not only $\ket{\psi_1}^s$, but all the  $\ket{\psi_N}^s$ states, could constitute a generalized W-class state, similar to the one proposed by Kim and Sanders~\cite{KyS}. In which follows we ellucidate this question by evaluating the monogamy relation for the $\ket{\psi_N}^s$ states.

\subsection{Monogamy relation of the $\ket{\psi_N}^s$ states}

For multipartite qubit systems, the monogamy relation of entanglement provides a way to characterize the different types of entanglement distribution. In other words, it relates the amount of entanglement between any two parties to the entanglement between those parties and the others. In 3-qubit systems, the monogamy relation corresponds to the Coffman-Kundu-Wootters (CKW) inequality~\cite{CKW}, given by
\begin{equation}\label{eq:CKW}
	{\mathcal C}^2_{A(BC)}\geq {\mathcal C}^2_{AB} + {\mathcal C}^2_{AC}\,.
\end{equation}

Here, ${\mathcal C}_{A(BC)}$ is the concurrence of a tripartite state between subsystem $A$ and the pair of subsystems $BC$. Terms at the right side of the inequality are concurrences of the reduced (generally mixed) states $\rho_{AB}$ and $\rho_{AC}$. Inequality \eqref{eq:CKW} is saturated by W-states, $\ket{W}=c_1\ket{100}+c_2\ket{010}+c_3\ket{001}$, with $c_i\in\mathbb C$, implying that the genuine tripartite entanglement of these states is completely determined by their partial entanglements between pairs $A-B$ and $A-C$.  A generalization of the previous definitions and results into $n$-qubit systems has been obtained~\cite{KyS,CKWmulti}.

For qudit systems, Inequality \eqref{eq:CKW} is no longer valid; 
counter examples can be found by extending the dimension of any of the subsystems. However, it has been shown that generalized W-states satisfy the same equality that in the qubit case, ${\mathcal C}^2_{A(BC)}= {\mathcal C}^2_{AB} + {\mathcal C}^2_{AC}$,~\cite{KyS}. We now evaluate both terms in the monogamy inequality for the non-diffractive eigenstates of the stub lattice. First, consider the term ${\mathcal C}^2_{A(BC)}=2(1-{\rm tr}(\rho_A^{(N)})^2)$, where $\rho_A^{(N)}=\rm tr_{BC}\left\lbrace\ket{\psi_N}^s\bra{\psi_N}\right\rbrace$ (all the results below are independent of the particular choice for the pairs of sites). From the general definition of states $\ket{\psi_N}^s$, Eq.\thinspace\eqref{eq:psins}, we obtain that the reduced density matrix is diagonal, $\rho_A^{(N)}=\left(\sum_{M=0}^N\binom{N}{M}2^{N-M}\ket{M}\bra{M}\right)/3^N$, which simplifies the expression for the concurrence between $A$ and $BC$:
\begin{equation}\label{eq:c2ABC}
 {\mathcal C}^2_{A(BC)}=2-\frac{2}{3}\sum_{M=0}^N\binom{N}{M}^2 2^{2(N-M)}\ .
\end{equation}
 
 Now, concurrence  ${\mathcal C}_{AB}$ (or ${\mathcal C}_{AC}$) is not as easy to compute, since the corresponding state $\rho_{AB}^{(N)}$ (or $\rho_{AC}^{(N)}$) is a mixed state. Then, its value is given by the minimum average concurrence taken over all the possible pure state decompositions of the density matrices.  Again, the general definition of  localized eigenstates $\ket{\psi_N}^s$ (and $\ket{\psi_N}^r$)  provides a convenient form  for the reduced matrix $\rho_{AB}={\rm tr}_{C}\ket{\psi_N}^s\bra{\psi_N}$, which can be expressed as a block matrix in terms of the non-diffractive states of the rhomboidal lattice. For the $\rho_{AB}^{(N)}$ state, we obtain

\begin{widetext}
\newcommand{\matindex}[1]{#1}
\begin{equation}
\rho_{AB}^{(N)}=
 \begin{blockarray}{cc|c|c|c}
 &  \matindex{\bra{00}} &
  \begin{array}{cc}
   \matindex{\bra{10}} & \hspace*{1em}\matindex{\bra{01}}
  \end{array} & 
  \begin{array}{ccc}
\matindex{\bra{11}} \hspace*{2em} & \matindex{\bra{20}} & \hspace*{2em} \matindex{\bra{02}}
\end{array} & 
 \\ 
 \vspace{-.5em} & & & & \\
\begin{block}{c(c|c|c|c)}
\matindex{\ket{00}} & 1 & 0  & 0 & \cdots
\\ \vspace{-.75em} & & & & \\ \BAhline \vspace{-.75em} & & & & \\ 
\begin{array}{c}
\matindex{\ket{10}} \\ \vspace{-.5em} \\ \matindex{\ket{01}} 
\end{array} &
0 &
\begin{array}{rr}
 \binom{N}{1} & -\binom{N}{1} \\ \vspace*{-.5em} & \\ -\binom{N}{1} & \binom{N}{1}
\end{array}
& 0 & \cdots
\\ \vspace{-.75em} & & & & \\ \BAhline \vspace{-.75em} & & & & \\ 
\begin{array}{c}
\matindex{\ket{11}} \\ \vspace{-.5em} \\ \matindex{\ket{20}} \\ \vspace{-.5em} \\ \matindex{\ket{02}} 
\end{array} &
0 & 0 &
\begin{array}{rrr}
2\binom{N}{2} & \sqrt{2}\binom{N}{2} & -\sqrt{2}\binom{N}{2} \\ \vspace*{-.5em} \\ \sqrt{2}\binom{N}{2} & \binom{N}{2} & \binom{N}{2}\\ \vspace*{-.5em} \\ -\sqrt{2}\binom{N}{2} & \binom{N}{2} & \binom{N}{2}
\end{array}
& \cdots
 \\ \BAhline \vspace{-.75em} & & & & \\ & \vdots & \vdots & \vdots & \ddots \\
\end{block}
\end{blockarray}
\;=\frac{1}{3^N}\sum_{M=0}^N 2^M\binom{M}{N}\ket{\psi_M}^r\bra{\psi_M}\ .
\end{equation}
\end{widetext}

Since all the $\ket{\psi_M}^r$ states are linearly independent, the Hughston-Jozsa-Wootters theorem allow us to obtain any other pure-state decomposition of $\rho_{AB}^{(N)}$, say $\sum_k^r \ket{\tilde \phi_k}_{AB}\bra{\tilde \phi_k}$ (with $r > N$), by operating on the set  of states $(2^M\binom{M}{N}/3^N)^{1/2}\ket{\psi_M}^r$ with an $r\times r$ unitary matrix. Consequently, we find that the squared average concurrence ${\mathcal C}_{AB}^2$ remains the same for any decomposition of $\rho_{AB}^{(N)}$, and it is given by
\begin{equation}\label{eq:c2AB}
\begin{split}
 {\mathcal C}_{AB}^2=\left[\frac{1}{3^N}\sum_{M=0}^N 2^M\binom{M}{N}{\mathcal C}(\ket{\psi_M}^r)\right]^2\ .
 \end{split}
\end{equation}

Notice that this result is the same as ${\mathcal C}_{AC}^2$, since the only difference is given by the powers of $-1$ associated with site $B$, which do not change the Schmidt coefficients nor the concurrence of  states contributing to the average. Comparation of ${\mathcal C}_{A(BC)}^2$ in Eq.\thinspace\eqref{eq:c2ABC} with ${\mathcal C}_{AB}^2={\mathcal C}_{AC}^2$ in Eq.\thinspace\eqref{eq:c2AB}, it can be seen that localized eigenstates of the stub lattice do not satisfy the monogamy relation characteristic of generalized $W$-states. Thus, although entanglement remains after remotion (tracing) of one site, it does not determine the amount of entanglement of the full tripartite state.

\section{Application to multi-core optical fibers}

The need to increase the data-carrying capacity of single optical fibers has led to commercial systems which currently employ multiplexing in frequency, time, polarization and phase. Only recently, space-division multiplexing (SDM) has become feasible~\cite{Richardson,Saitoh}. This method enables data transmission through several physically distinguishable propagation paths, as for instance, in the case of multicore fibers (MCF)~\cite{MCF}. Here, the paths are defined by an array of single-mode cores within a single fiber. Susceptibility to fractures currently limits MCF to a diameter smaller than 230 $\mu$m~\cite{Matsuo}. Crosstalk, that is, photon hopping between nearest-neighbor cores, and photon losses are the primary sources of error. 

Crosstalk can be greatly decreased by decoupling the cores; that is, by placing the single-mode cores well separated in such a way that the coupling constant becomes very small. This strategy, however, constraints the maximal number of cores within a single fiber. Currently, a 200 $\mu$m outer diameter MCF with 19 cores has been reported~\cite{19MCF}. This reached an effective transmission distance of approximately 10 km. In order to increase the core density several approaches have been proposed. Heterogeneous multi-core fibers~\cite{HMCF}, composed by single-mode cores with different propagation constants, allow to limit crosstalk while reducing the distance between cores. Small random variations in core properties also lead to a strong crosstalk suppression~\cite{RMCF}. Fiber bend and trench-assisted cores have led to ultra-low crosstalk MCF~\cite{BMCF}.

\begin{figure}[]
	\includegraphics[width=.48\textwidth]{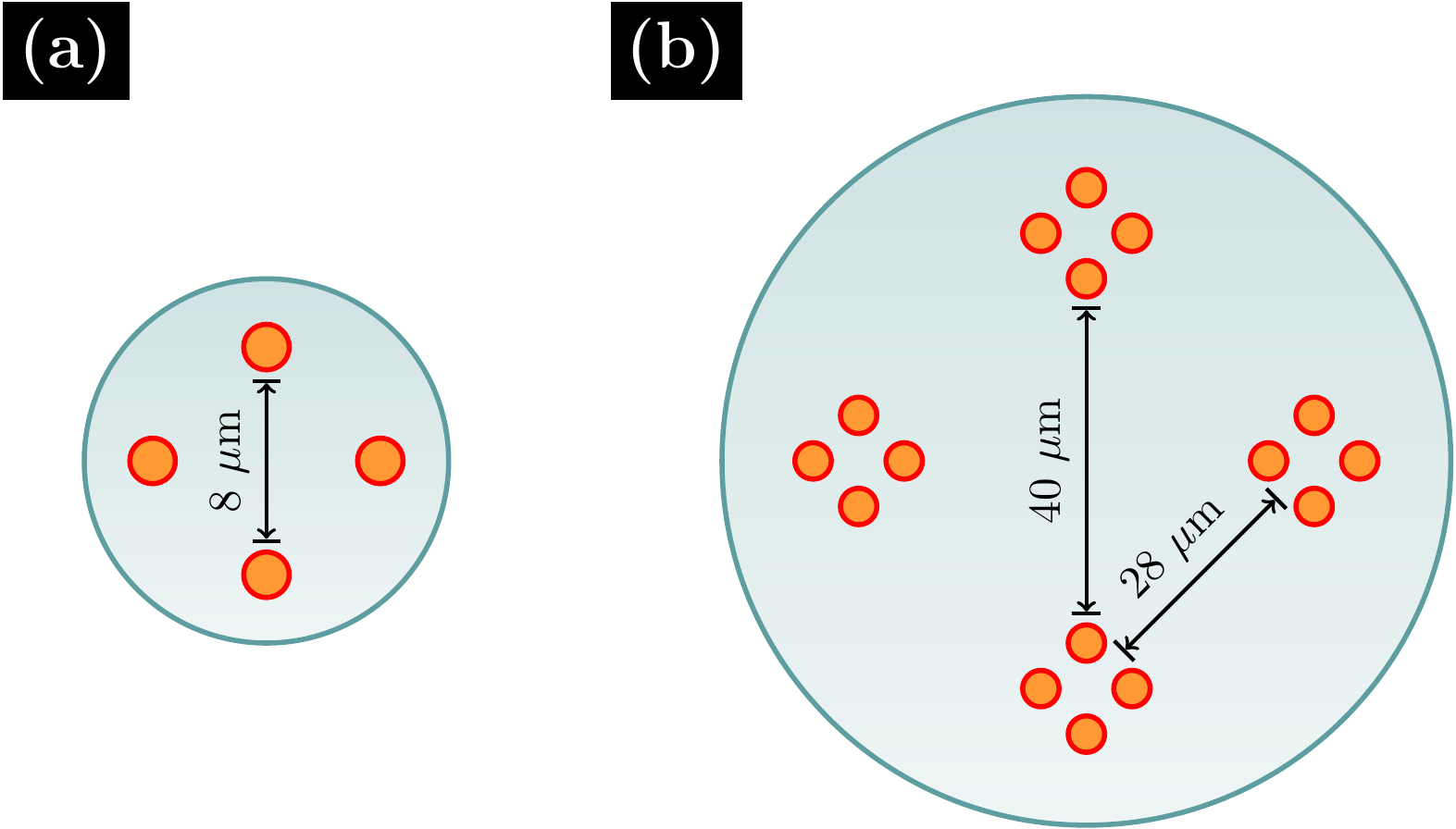}
	\caption{Designs of multi-core optical fibers sustaining non-diffractive modes. (a) Four-core fiber, where two modes can be transmitted. (b) A more complex design, composed of four  cells with four cores each. Distances shown correspond to realistic separation between cores allowing suppression of cross-talk.}\label{fg:fibers}
\end{figure}

Localized quantum states of light offer a new approach to suppress crosstalk. To illustrate this, let us consider a recently reported experiment~\cite{Canas} where a single MCF has been exploited for the experimental realization of quantum key distribution via high-dimensional quantum systems. Here, a single photon encoding a four-dimensional quantum state is transmitted through a single-mode fiber with four cores arranged in a diamond-like cross-section, like the four sites illustrated in Fig.\thinspace\ref{fg:fibers} (a). In this setup we can resort to localized quantum states of light to completely suppress crosstalk independently of the coupling constant between waveguides. We can define the separable states
\begin{equation}
|\sqrt{2}\alpha\rangle^r_{A,B}=|\alpha\rangle_A|-\alpha\rangle_B|0\rangle_{S'}|0\rangle_{S}
\end{equation}
and
\begin{equation}
|\sqrt{2}\alpha'\rangle^r_{-1,+1}=|0\rangle_A|0\rangle_B|\alpha'\rangle_{S'}|-\alpha'\rangle_{S}\ ,
\end{equation}
which remain localized in cores $A$ and $B$ and $S'$ and $S$ (as in Fig. \ref{fg:geom}(a)), respectively, along the propagation in the MCF. Each one of these two states allows the encoding and transmission of information employing two cores of the MCF. Thus, two crosstalk-free communication channels can be generated in a four-core MCF. In order to increase the number of effective communication channels we can form four groups each consisting of four cores, as illustrated in Fig.\thinspace\ref{fg:fibers}(b).  Here, cores 1 to 4 have the same geometry and distances as in Fig.\thinspace\ref{fg:fibers}(a), while cores in each group can be very close without exhibiting crosstalk. This is due to the fact that even if all four cores are coupled, the previous states are still localized states of light. Thereby, the crosstalk level, given by cores 1 to 4, is equivalent to a four-core MCF while establishing 8 communication channels with 16 cores. An hexagonal ring of four-core groups would allow encoding and transmission through 12 effective channels with 24 cores.

This strategy to generate crosstalk-free channels is compatible with other proposals having the same goal. For instance, we can consider a MCF formed by heterogeneous four-core groups. All cores in a group are homogeneous, that is, they have the same propagation constant $\epsilon$, but a different value in each group. Heterogeneous MCFs leads to an important reduction of the crosstalk as well as to a denser packing of cores, when compared to homogeneous MCFs. A further reduction can be obtained by resorting to trench-assisted cores, which decrease the coupling constant between neighboring cores by carefully engineering the index refraction profile of each core.

Our strategy it is not restricted to separable states. We can consider single-photon path-entangled states such as
\begin{equation}
|\psi_1\rangle^{r}_{A,B}=\frac{1}{\sqrt{2}}(|1\rangle_A|0\rangle_{B}-|0\rangle_A|1\rangle_{B})|0\rangle_{S'}|0\rangle_{S}
\end{equation}
and
\begin{equation}
|\psi_1\rangle^{r}_{-1,+1}=\frac{1}{\sqrt{2}}|0\rangle_A|0\rangle_B(|1\rangle_{-1}|0\rangle_{+1}-|0\rangle_{S'}|1\rangle_{S})\ .
\end{equation}
These two orthogonal states remain localized in their respective sites along the propagation in the MCF. Thereby, any arbitrary superposition of states $|\psi_1\rangle^{r}_{A,B}$ and $|\psi_1\rangle^{r}_{S',S}$ will be coherently preserved. This encoding make thus possible the transmission of a single qubit. Similarly, states of the form $\alpha|\psi_N\rangle^r_{A,B}+\beta|\psi_M\rangle^r_{S',S}$ with arbitrary total photon numbers $N$ and $M$ are also not affected by crosstalk. This shows that MCFs are promising candidates to reliably distribute path-entangled states among several parties. 

A second important source of errors is the fast absorption of photons by the cladding of optical fibers, that is, photon losses. In the case of a single-core optical fiber this process can be modeled by the interaction Hamiltonian $H_{loss}=\bar\kappa(ab^\dagger+a^\dagger b)$ in the Schr\"odinger picture, where $a$ and $b$ are the annihilation operators of photons at core and cladding, respectively, and $\bar\kappa$ is the coupling constant. This Hamiltonian leads to the Master equation for the state of photons propagating in the core
\begin{equation}
\frac{d\rho}{dt}=\frac{\kappa'}{2}(a^\dagger a\rho+\rho a^\dagger a-2a\rho a^\dagger)\ ,
\label{ME}
\end{equation}
which describes the progressive loss of energy from the core to a zero-temperature cladding (environment), that is, $\langle b^\dagger b\rangle=0$. This model is known as amplitude damping~\cite{Louisell,Gardiner}. Here, $\kappa'=2\bar\kappa^2\eta\tau_c$, where $\eta$ is a constant arising at the integration of Heisenberg equation for both core and cladding and $\tau_c$ is the correlation time of the cladding. The state of the core generated by the Master equation (\ref{ME}) after a time interval $\delta t$ can be cast in the form~\cite{Chuang}
\begin{equation}
\rho({\delta t})=\sum_kE_k\rho(0)E_k^\dagger\ ,
\end{equation}
with
\begin{equation}
E_k=\sqrt{\frac{(1-e^{-\kappa'\delta t})^k}{k!}}e^{-\kappa'\delta t a^\dagger a}a^k\ .
\end{equation}
Considering the case of a two-core MCF, the state of the field at the core after a time interval $\delta t$ can be cast as
\begin{equation}
\rho({\delta t})_{A,B}=\sum_{k,m}E_m^{B}E_k^{A}\rho(0)_{A,B}(E_k^A)^\dagger(E_m^B)^\dagger\ ,
\end{equation}
where we have assumed a common environment (the cladding) for both cores. The action of the annihilation operators $a_A$ and $a_B$ onto state
$|\psi_N\rangle^{r}_{A,B}$ is
\begin{equation}
a_A|\psi_N\rangle^{r}_{A,B}=\frac{1}{\sqrt{2}}|\psi_{N-1}\rangle^{r}_{A,B}
\end{equation}
and
\begin{equation}
a_B|\psi_N\rangle^{r}_{A,B}=-\frac{1}{\sqrt{2}}|\psi_{N-1}\rangle^{r}_{A,B}\ .
\end{equation}
This indicates that even after the loss of a single photon, light remains in a localized quantum state. Thereby, if the initial state $\rho(0)_{A,B}$ of both cores A and B is given by $|\psi_N\rangle^{r}_{A,B}$, then the state $\rho(\delta t)_{A,B}$ becomes an incoherent convex combination of all mutually orthogonal localized quantum states of light with total photon number less or equal than $N$. It is thus possible to encode a qubit as $\alpha|\psi_N\rangle^{r}_{A,B}+\beta|\psi_{N+M}\rangle^{r}_{A,B}$ with $M\ge1$. After the loss of $N-1$ photons the coherence of the state is still preserved as $\alpha|\psi_1\rangle^{r}_{A,B}+\beta|\psi_{1+M}\rangle^{r}_{A,B}$.

\begin{figure*}[!tb]
	\centering
	\includegraphics[width=.95\textwidth]{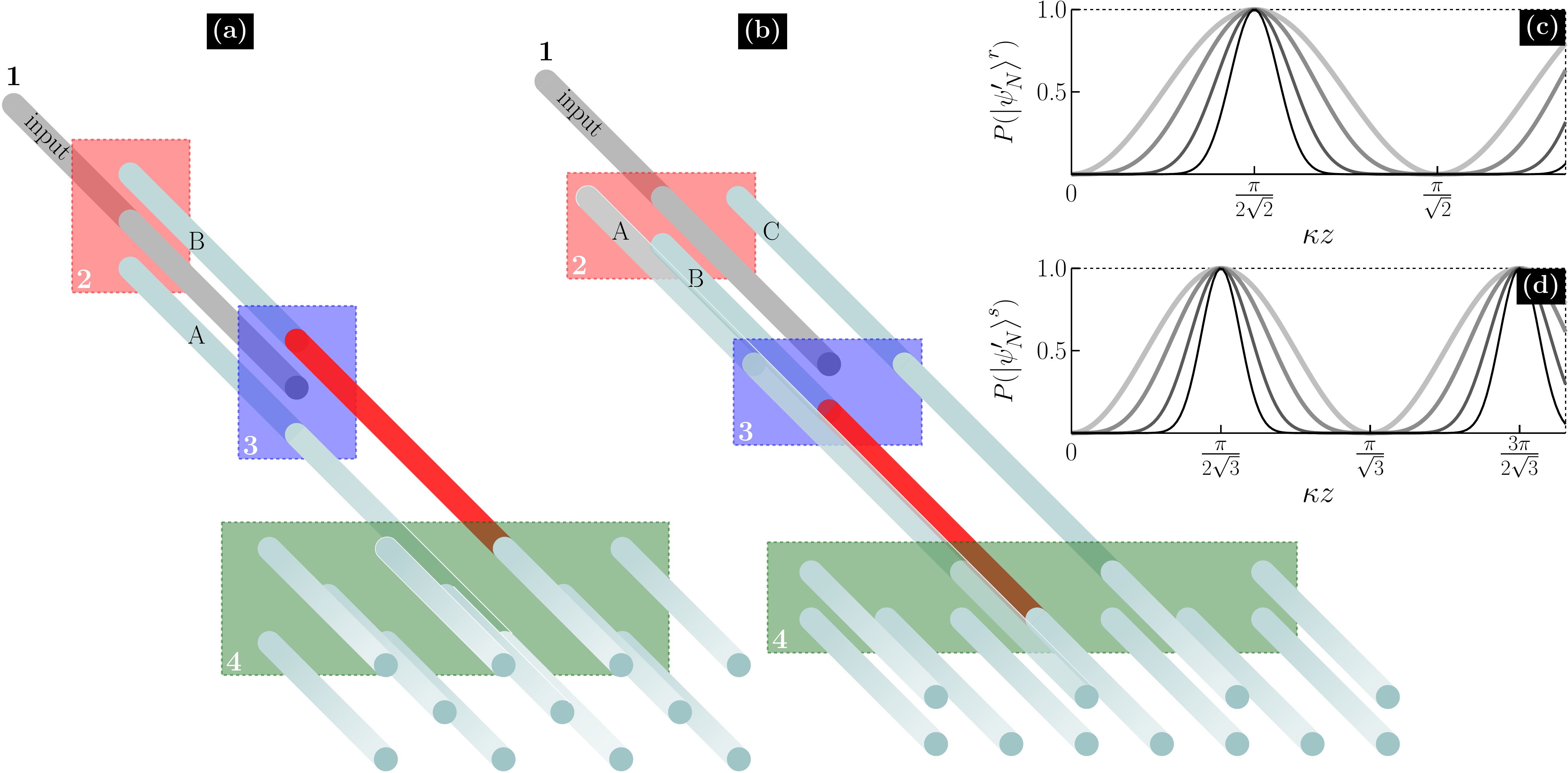}
	\caption{Scheme for the preparation of localized quantum states of light in the (a) rhomboidal lattice and in the (b) stub lattice. In both cases, a Fock state $\ket{N}$ coupled to an input waveguide at stage 1, is converted in a localized eigenstate $\ket{\psi_N}^r$ or $\ket{\psi_N}^s$, respectively, and finally injected at certain position of the array at stage 4. (c) Probability of finding the state $\ket{\psi_N'}^r$ on waveguides $A$ and $B$ (equal to the state $\ket{\psi_N}^r$ but without the $(-1)^q$ factors) as a function of distance between stages 2 and 3 of setup (a). (d) The same as (c) for states $\ket{\psi_N'}^s$ found in waveguides $A$, $B$ and $C$ between stages 2 and 3 of setup (b). Lines in (c) and (d), from the thickest to the thinest, correspond to $N=1$, $N=2$, $N=5$ and $N=12$.}\label{fg:props}
\end{figure*}

\section{On-chip preparation of localized quantum states of light}

All the localized eigenstates $\ket{\psi_N}^r$ and $\ket{\psi_N}^s$ can be obtained by means of unitary linear operations over the corresponding Fock state $\ket{N}$. However, in practice it is very difficult to perform these operations with bulk optics, since it requires to control the phases of different multiphoton states distributed over different paths. Fortunately, it is possible to exploit the evanescent coupling between nearby waveguides to prepare the states for any number of photons.

In Fig.\thinspace\ref{fg:props}(a) and (b) we present two feasible designs, which respectively produce the localized eigenstates of the rhomboidal and the stub lattice at specific sites.  In both cases, a Fock state $\ket{N}$ is coupled to a single input waveguide at the point 1. From 2, photons tunnel to the neighboring waveguides via evanescent coupling of the field modes. The propagation of the photonic state in this section is described by a Hamiltonian of the same form as \eqref{eq:H}. If the waveguides are equally separated, after certain distance (stage 3) photons are found only on waveguides $A$ and $B$ for the diamond lattice, and in $A$, $B$ and $C$ for the stub one, forming a state $\ket{\psi_N'}$ which just differs from the desired state $\ket{\psi_N}$  in the absence of the $(-1)^q$ factors (phase structure). As depicted in Fig.\thinspace\ref{fg:props} (c) and (d), we find that the minimal propagation distance required to produce a state $\ket{\psi_N'}^r$ and $\ket{\psi_N'}^s$ is given by $l_c=\pi/2\sqrt{2}$ and $\pi/2\sqrt{3}$ respectively, and it is the same for any number of photons \cite{nota}
. In experiments with waveguides arrays, the value of the coupling constant is on the order of $0.5\,{\rm mm}^{-1}$ when waveguides are separated by $8\,\mu$m, implying that the length between stages 2 and 3 should be $\sim 5\,$mm, and can be varied in a wide range by changing the separation between the waveguides (due to the exponential decay of the coupling constant with the separation~\cite{Crespi,AldoSlim}). From this point, the input waveguide is suppressed, and, in order to introduce a phase factor $(-1)^q$ in the coefficients, a difference in the refractive index of one waveguide must be induced between stages 3 and 4 (red waveguide in Fig.\thinspace\ref{fg:props}), either by changing the writing speed in the fabrication process or by the application of an external electric field. This adds an increase equal to $\Delta \beta$ to the propagation constant on the chosen waveguide (this technique has been used in schemes which require a controlled phase shift, like the one in Ref.~\cite{2ndord}). The hamiltonian which rules the evolution in this section is given by $-\Delta \beta \hat{a}_{i}^\dagger \hat{a}_i$. This way, all the components of a state $\ket{\psi}$ expressed in the Fock base will acquire a phase depending solely on the number of photons in the waveguide with the $\Delta \beta$ increase. Then, we only need to find the length $\ell$ between stages 3 and 4 which gives a phase $\pi$ to a state with one photon in the chosen waveguide. This ensures that all the terms with an odd number of photons will acquire the same phase, but the terms with an even number of photons will not be affected. In other words, this section of the setup provides the $(-1)^q$ factor present in the definitions of the localized eigenstates \eqref{eq:psinr} and \eqref{eq:psins}. The desired phase shift could be achieved with a distance of $\sim 1$\,cm~\cite{2ndord}. Finally, at the point 4, the state is already prepared and can be put as initial condition to propagate along the array. Notice that althought the position in the lattice at which the state is injected must be fixed in the fabrication process, it is possible to construct input ports for several different positions in the same array, allowing the construction of more complex profiles and the implementation of protocols for image transmission
. Since the method that we propose here is completely {\it on-chip}, it can improve the reliability of all the applications which require precise combinations of the localized modes.

\section{Conclusion}

Here, we have studied the problem of localization of quantum light in flat-band lattices. In particular, we have considered rhomboidal, symmetric rhomboidal, stub, kagome, and Lieb lattices. We have constructed quantum states of light with a well defined photon number that stay localized along the propagation in the lattice, that is, photons described by these states do not exhibit diffraction. The localization is perfect, that is, photons always propagate along the same few initial sites, and is independent of external parameters. The localized quantum states of light are eigenstates of the interaction Hamiltonian with vanishing eigenvalue. This requirement together with the symmetry of the lattices leads to a vanishing probability for the hopping of photons from the initial sites to neighboring sites, which arises due to a destructive interference effect between hopping processes from different initial sites to a given neighboring site. In the basis of Fock states these are given by real (positive and negative) probability amplitudes corresponding to the square root of multinomial coefficients. Single-mode states with binomial probability distribution have been discussed previously in the literature~\cite{Aharonov,Stoler,Datolli}. A prominent feature of this class of states is that Fock and coherent states can be recovered as limit cases. A generalization of the single-mode binomial state to the multi-mode case, the so called multinomial states, has also been proposed~\cite{Gilmore,Fu}. The localized quantum states of light, Eq.\thinspace(\ref{eq:psinr}), belong to this latter class of states. Thereby, states $\ket{\psi_N}^r_{A,B}$ are $su(2)$ coherent states for any value of the total photon number $N$. 

We have also studied the entanglement properties of localized quantum states of light. In the bipartite case of the rhomboidal lattice, whose localized states $|\psi_N\rangle^r_{A,B}$ involve two sites only, states are entangled for any value of the total photon number $N$. In the tripartite case of the stub lattice, whose localized states $|\psi_N\rangle^s_{A,B,C}$ involve three sites, any bipartite partition of the three sites is entangled for any value of the total photon number $N$. Tracing out photons propagating at a particular site generates a new bipartite reduced mixed state which is also entangled. Furthermore, it was possible to show that the localized states $|\psi_N\rangle^r_{A,B,C}$ do not satisfy the monogamy relation of Coffman-Kundu-Wootters and consequently do not correspond to generalized W-states.

An interesting application of localized quantum states of light arises in the context of a multicore fiber, where the propagation paths for the light are defined by an array of single-mode cores within a single fiber. Two important source of errors, which limit the effective transmission length, are photon crosstalk and photon losses. The former can be passively suppressed by resorting to localized quantum states of light of a given lattice. The latter reduces the total photon number of the localized state without destroying the localization property of the state. Let us note that this result shows the possibility of reliably transmitting multi-photon path-entangled states through multicore fibers.

We have proposed a setup for the experimental generation of localized quantum states of light. This setup requires the ability to generate arbitrary Fock states and all operations are carried inside a photonic crystal. Experimental demonstrations of localized quantum states of light are within reach of current experimental capabilities for 1 and 2 photons.

Our results can be extended in several ways. Elliptical femtosecond-laser-written waveguide arrays exhibit an asymmetry of the spatial transverse profiles of linearly polarized modes in these waveguides. This leads to a polarization-dependent coupling coefficient $\kappa$ between adjacent waveguides that strongly affects the propagation of light on a lattice~\cite{AldoSlim}. Since the localized quantum states of light do not depend on the value of the coupling constant, we obtain a set of this class of states for each orthogonal polarization. The application to MCF has been discussed considering several cells each one composed of four cores. This choice is motivated by recent experiments on the propagation of single photons that employed four-core MCF. It is possible, however, to envisage more complex lattices. For instance, Lieb or Kagome lattices filling the complete area of the fiber, as in the case of hollow-fibers \cite{Poletti}, might lead to an increase in the ratio between communication channels and cores. Recently, a procedure has been reported that allows the construction of a large family of lattices with non-trivial geometries supporting one or more FBs~\cite{Luis} even in the presence of next-nearest-neighbor coupling. This opens new possibilities for the experimental implementation of FB quantum localized states using different geometrical configurations, depending on the particular setup, and to the possibility of employing larger numbers of localized quantum states of light in single and multi-mode MCF.

\acknowledgments

This work was supported by Millennium Scientific Initiative Grant No. RC130001 and FONDECyT Grants No. 1140635 and No. 1151444.


\begin{thebibliography}{99}
\bibitem{nota} Although the invariance of the coupling length with respect to the number of photons is expectable (because we are dealing with linear coupling only), it is not possible to show it explicitely, due to the lack of an analytical expression for the eigenstates of the hamiltonian in the general $N$-photon case.
\end{thebibliography}
\end{document}